\documentclass[prd,aps,floats,preprintnumbers,preprint]{revtex4}

\usepackage{graphicx}
 
\newcommand{\be}{\begin{equation}}
\newcommand{\ee}{\end{equation}}
\newcommand{\bea}{\begin{eqnarray}}
\newcommand{\eea}{\end{eqnarray}}

\newcommand{\PTP}{\it Prog. Theo. Phys.}

\newcommand{\PR}{{\it Phys. Rev.\,}}

\begin{document}

\setlength{\unitlength}{1mm}

\title{Discrete gravity from statistical mechanics}

\author{Antonio Enea Romano}
\affiliation{
$^1$Leung Center for Cosmology and Particle Astrophysics, National Taiwan University, Taipei 10617, Taiwan, R.O.C.\\
\\
}


\begin{abstract}
We show how to construct space time lattices with a Regge action proportional to the energy of a given Ising or Potts model macrostate. This allows to take advantage of the existence of exact solutions for these models to calculate the quantum wave function of the universe using the sum over the histories approach to quantum gravity.
Motivated by this isomorphism we show how the Regge equations, i.e. the discrete equivalent of the vacuum Einstein equations, can be derived using statistical mechanics under the assumption that the energy of a given space time geometry is proportional to the Regge action.

\end{abstract}

\maketitle

\section{Introduction}

There has been over time a considerable interest about gravity as an emergent theory \cite{Verlinde:2010hp,Sakharov:1967pk,Jacobson:1995ab}.
Here we will propose a discrete realization of such a conjecture, starting from a formal analogy between the Regge action and the Ising model energy. We will then generalize this and show how the discrete equivalent of the Einstein's equation can be derived from statistical mechanics by an appropriate definition of the energy of a given space time geometry.

Regge calculus is a discrete approach to gravity consisting in representing space time with a flat simplicial complex. It does not require the use of any coordinate system, and this will prove to be an important feature to draw the connection with statistical mechanics, since the fundamental object in terms of which to describe space time geometry configuration will not be the metric, but rather local curvature.
The key quantity to define the configurations of the spacetime lattice is the deficit angle which is computed, given
the edge lengths of the lattice, using
\begin{equation}
  \epsilon = 2\pi - \sum_k \theta_{k}
\end{equation}
where the summation is over all four-simplices $k$ which contain the
triangle, and $\theta_k$ is the hyper-dihedral angle between the two
tetrahedral faces which hinge on the triangle within $k$. 
The simplicial equivalent of the Hilbert action
\cite{regge61} for a lattice spacetime is,
\begin{equation}
  \int_{\cal M} \sqrt{-g}\, R\, d^4x \quad \rightarrow \quad 2 \sum
  _{i} \, \epsilon_i A_i=I^R
\end{equation}
where the summation is over all triangles $i$, $A_i$ is the area and
$\epsilon_i$ is the deficit angle of the $i$th triangle. 
Lattice edges are the discrete equivalent of the metric, and can be used to 
construct any other geometric quantity.  In its original formulation by Regge the equations of motion are obtained by a variational principle; the action is varied with
respect to the independent variables, i.e. the edges lengths.  This yields the vacuum ``Regge
equations''
\begin{equation}
  0 = \frac{\delta I^R}{\delta l_j} = 2 \sum_i \epsilon_i \frac{\delta
  A_i}{\delta l_j}.
\end{equation}
In deriving the second equality in the equation above it has been used the Regge identity
\cite{regge61},
\begin{equation}
  \sum_i \frac{\delta \epsilon_i}{\delta l_j} A_i = 0. 
\end{equation}
which is the discrete equivalent of the Palatini identity.
A good review of classical applications of Regge calculus can be found for  in \cite{Gentle:2002ux}, and in the context of quantum gravity in \cite{williams97}. 

\section{Mapping Ising and Potts models into simplicial space time lattices}
Ising models are simplified mathematical models studied in statistical mechanics, consisting in a set variables $S_i$, called spin, which can only take values 1 or -1.
The energy for one dimensional Ising models is given by:
\be
E^{I} =-J \sum_{i}^N S_{i} S_{i+1}=\sum_i^{N-1} E^{I}_i .\label{EI}
\ee
For the purpose to find a connection with statistical mechanics we can now rewrite the Regge action in terms its local contribution $I^R_i=2 \epsilon_i A_i$ according to:
\be
I^{R}=2 \sum_{i}^{N-1} \, \epsilon_i A_i=\sum_{i}^{N-1} I^{R}_i, \label{IR}
\ee
where $I^{R}_i$ can be interpreted as the discrete equivalent of the $R\, dV$ in the Hilbert action.

By comparing the two expressions (\ref{EI},\ref{IR}) we see that for a given Ising model macrostate determined by its spin configuration we can construct a corresponding space time lattice according the prescription that:
\be
E^{I}_i=E^{R}_i=\alpha I^R_i ,\,\,\, 1\leq i \leq N-1,
\ee
where we have introduced a constant $\alpha$ for dimensional consistency, 
and $E^{R}_i$ can be interpreted as the local contribution to the total energy of a given space time configuration.

In Ising models there are only two possible values of $E^{I}_i$, so not all possible space time geometries can be mapped into a Ising model and vice versa, and the the corresponding space time lattice would be some kind of "digital" perturbation of flat space time, i.e. with only two possible value of $I^R_i$, corresponding to local positive or negative curvature.
A similar correspondence could easily be extended to one dimensional Potts models, for which $E_i$ can instead take an arbitrary number of values. 
In this models in fact the spin take one of q possible values, distributed uniformly about the circle, at angles:
\be
\Theta_n = \frac {2 \pi n} {q},
\ee
and in the one dimensional case the Hamiltonian is similar to the Ising model :
\be
E^P  =-J \sum_{i}^N S_{i}\cdot S_{i+1}=\sum_i^{N-1} E^{P}_i,
\ee
where now the scalar product is given by:
\be
S_{i}\cdot S_{j}=\cos \left (\theta_i {} - \theta_j {} \right).
\ee
Since exact solutions are known for the one dimensional Ising and Potts models, the corresponding partition function can be used to study the wave function of the universe \cite{Hartle:1983ai,Hartle:1986up,Hartle:1985wr} using the sum over the histories approach to quantum gravity.
In the case of Potts models there is more freedom in the number of possible local values of  $E^R_i=E^P_i$ , but there is still a bound given by J:
\be
|E_i|\leq J
\ee
which can be interpreted as the expected maximum amplitude of the quantum fluctuations of the space time geometry.
The analogy we have shown so far suggests that the fundamental quantity which needs to be quantized is not the metric but rather the Hilbert action integrand $R\, dV \approx I^{R}_i$, i.e. the curvature times the four dimensional volume element. In other words space time configurations could be described by an appropriate set of $I^R_i$ without the need of introducing any coordinate system or metric.  

\section{Deriving discrete gravity from statistical mechanics}
Inspired by the results of the previous section we can go further and assume in general that
any space time geometry, when approximated by a simplicial lattice, has an energy given by :
\be
E^R=\sum_i E^R_i.
\ee
Under this assumption we can try to derive the Regge equations not from an action principle, which was the original way Regge derived the vacuum equations, but rather using a statistical mechanics approach.

First we need to introduce some notation to define different macrostates :
\be
I^{R,j},
\ee
where different values of the upper index 'j' correspond to different lattice configurations, i.e. different set of values of $I^R_i$.
After introducing the partition function:
\bea
Z=\sum_j e^{-\beta E^{R,j}},
\eea
we can then define the probability of a given space time lattice configuration as:
\bea
P_j=\frac{e^{-\beta E^{R,j}}}{Z}.
\eea
If we now look for the lattice configuration with maximum probability we get:
\bea
\frac{\partial E^{R,Min}}{\partial l_i}=0,
\eea
which are clearly equivalent to the Regge vacuum equations and we have maximized respect to the the lattice bones lengths $l_i$.
In this way we have been able to derive the discrete simplicial equivalent of the vacuum Einstein equations, i.e. the Regge equations, assuming only statistical mechanics and interpreting the Regge action as the key quantity to define the energy of a given space time lattice configuration.
Under this interpretation different space time configurations have different energy, and the the most probable one, i.e. the one with the lowest energy and maximum entropy, corresponds to the same configuration predicted by the discrete equivalent of general relativity, i.e. the Regge equations.

\section{Conclusions}

In this paper we have first shown some formal analogy between the Regge action and Ising and Potts models energy, and inspired by it we have proposed how to construct space time lattices having the same partition function. 
This allows to take advantage of the existence of exact solutions for Ising models, to study the quantum wave function of the corresponding simplicial geometries.
We have then generalized this analogy, and proposed that any space time configuration has a local energy which is proportional to the curvature times the four dimensional volume element, and that in the case of a simplicial lattice is proportional to the local contribution $I^R_i$ Regge action $I^R$.Under this assumptions and using only statistical mechanics we have derived the discrete equivalent of the vacuum Einstein equations, i.e. Regge equations, which correspond to a constraint which must be satisfied by the most probable space time geometry, which is the one with the lowest energy.

It will be interesting in the future to find a continuum equivalent of the present discrete approach to derive gravity from statistical mechanics, and find other statistical mechanics systems which are well understood theoretically and can be mapped into some isomorphic space time lattice configurations.

\begin{acknowledgments}
Romano is supported by the Taiwan NSC under Project No.\
NSC97-2112-M-002-026-MY3, by Taiwan's National Center for
Theoretical Sciences (NCTS).

\end{acknowledgments}

\end{document}